\documentclass[aps,prb,twocolumn,showpacs,superscriptaddress,floatfix]{revtex4-1}

\usepackage{graphicx}
\usepackage{bm}
\usepackage[usenames]{color}
\usepackage[breaklinks=true,colorlinks,citecolor=blue,linkcolor=blue,urlcolor=blue]{hyperref}
\usepackage[abs]{overpic}
\usepackage{ulem}

\newcommand{\xs}{\tilde{x}}
\newcommand{\ys}{\tilde{y}}
\newcommand{\ks}{\tilde{k}}
\newcommand{\ft}{\hat}

\begin{document}

\title{Non-local transport and the hydrodynamic shear viscosity in graphene}
\author{Iacopo Torre}
\email{iacopo.torre@sns.it}
\affiliation{NEST, Scuola Normale Superiore, I-56126 Pisa,~Italy}
\affiliation{Istituto Italiano di Tecnologia, Graphene Labs, Via Morego 30, I-16163 Genova,~Italy}
\author{Andrea Tomadin}
\affiliation{NEST, Istituto Nanoscienze-CNR and Scuola Normale Superiore, I-56126 Pisa,~Italy}
\author{Andre K. Geim}
\affiliation{School of Physics \& Astronomy, University of Manchester, Oxford Road, Manchester M13 9PL, United Kingdom}
\author{Marco Polini}
\affiliation{Istituto Italiano di Tecnologia, Graphene Labs, Via Morego 30, I-16163 Genova,~Italy}
\affiliation{NEST, Scuola Normale Superiore, I-56126 Pisa,~Italy}

\begin{abstract}
Motivated by recent experimental progress in preparing encapsulated graphene sheets with ultra-high mobilities up to room temperature, we present a theoretical study of dc transport in doped graphene in the hydrodynamic regime. By using the continuity and Navier-Stokes equations, we demonstrate analytically that measurements of non-local resistances in multi-terminal Hall bar devices can be used to extract the hydrodynamic shear viscosity of the two-dimensional (2D) electron liquid in graphene. We also discuss how to probe the viscosity-dominated hydrodynamic transport regime by scanning probe potentiometry and magnetometry. Our approach enables measurements of the viscosity of any 2D electron liquid in the hydrodynamic transport regime.
\end{abstract}

\maketitle

\section{Introduction}
\label{sect:intro}

Transport in systems with many particles (such as gases and liquids) undergoing very frequent inter-particle collisions has been studied for more than two centuries and is described by the theory of hydrodynamics~\cite{landaufluidmechanics,batchelor,huangbook}. In the hydrodynamic regime, transport is described by~\cite{landaufluidmechanics,batchelor,huangbook} three non-linear partial-differential equations---the continuity, Navier-Stokes, and energy-transport equations---reflecting the conservation of mass, momentum and energy, respectively. The Navier-Stokes equation contains two transport coefficients~\cite{landaufluidmechanics,batchelor,huangbook}: the shear viscosity, $\eta$, which describes friction between adjacent layers of fluid moving with different velocities, and the bulk viscosity, $\zeta$, which describes dissipation arising in a liquid due to homogeneous compression-like deformations.
The energy transport equation contains the thermal conductivity $\kappa$, which describes dissipative heat flow between regions with different temperatures. These coefficients quantify the tendency of the liquid to restore a homogeneous state in response to a velocity or thermal gradient: they therefore control the magnitude of non-local contributions to the linear-response functions of the liquid.

Viscous flow of dilute classical gases attracted the attention of Maxwell, who theoretically discovered~\cite{huangbook} a puzzling property of the shear viscosity of a dilute gas.
Using a molecular approach~\cite{huangbook}, he found that the shear viscosity of a dilute gas is independent of density $n$ (and depends on temperature $T$ according to $\eta \propto T^{1/2}$), a counterintuitive result that he felt needed immediate experimental testing~\cite{maxwell}. The importance of $\eta$ in the hydrodynamic behavior of dilute gases and liquids stems from the fact that this parameter controls the term of the Navier-Stokes equation that opposes turbulent flow~\cite{landaufluidmechanics,batchelor,huangbook}.

Recent years have witnessed a tremendous interdisciplinary interest in the hydrodynamic flow of strongly interacting quantum fluids.
This interest was sparked by a series of results~\cite{blackholes}, which were obtained via the anti-de Sitter/conformal field theory (AdS/CFT) correspondence, for the shear viscosity of a large class of strongly interacting thermal quantum field theories.
These efforts culminated in 2005 when it was conjectured~\cite{AdsCFTbound} that all quantum fluids obey the following universal lower bound: $\eta/s \geq \hbar/(4\pi k_{\rm B})$, where $s$ is the entropy density.
Note that this bound does not contain the speed of light, thereby explaining why the conjecture was extended also to non-relativistic quantum field theories.
Fluids that saturate this bound have been dubbed ``nearly perfect fluids" (NPFs)~\cite{NPFsreview}, i.e.~fluids that dissipate the smallest possible amount of energy and satisfy the laws of hydrodynamics at distances as short as the inter-particle spacing.
Currently, two laboratory systems come closest to saturating the AdS/CFT bound: i) the quark-gluon plasma~\cite{QGP}, which is created at Brookhaven's Relativistic Heavy Ion Collider and at CERN's Large Hadron Collider by bashing heavy (e.g.~gold and lead) ions together and ii) ultracold atomic Fermi gases~\cite{cao_science_2011,elliott_prl_2014} (such as $^{6}{\rm Li}$) close to a Feshbach resonance.
Although mathematical counterexamples have appeared in the literature~\cite{cremonini2011}, there are no known experimental violations of the AdS/CFT bound.

The present work is motivated by the following questions: Do electron liquids display hydrodynamic behavior? If so, how can it be experimentally proven that the electron system has entered the hydrodynamic transport regime? Once in the hydrodynamic regime, how can the shear viscosity of an electron liquid be measured in a solid-state device? Can an electron liquid in a solid-state device be a NPF?

Hydrodynamics has been used for a long time to describe transport of electrons in solid-state devices~\cite{gurzhi_spu_1968,dyakonov_prl_1993,govorov_prl_2004,sai_prl_2005,muller_prb_2008,bistritzer_prb_2009,muller_prl_2009,andreev_prl_2011,mendoza_prl_2011,svintsov_jap_2012,mendoza_scirep_2013,tomadin_prb_2013,tomadin_prl_2014,narozhny_prb_2015}.
However, since the (Bloch) momentum of an electron in a solid is a poorly conserved quantity due to collisions against impurities, phonons, and structural defects in the crystal, experimental signatures of hydrodynamic electron flow are expected only in ultra-clean crystals, at sufficiently low temperatures.
Second, electron-electron (e-e) interactions need to be sufficiently strong to ensure that the mean free path $\ell_{\rm ee}$ for e-e collisions is the shortest length scale in the problem, i.e.~$\ell_{\rm ee} \ll \ell, W, v_{\rm F}/\omega$.
Here, $v_{\rm F}$ is the Fermi velocity, $\ell$ is the mean free path for momentum-non-conserving collisions, $W$ is the sample size, and $\omega$ the frequency of the external perturbation.

Unfortunately, the low-temperature requirement (needed to mitigate the breakdown of momentum conservation in a solid-state environment) and the strong e-e interaction requirement are conflicting: at low temperatures, where $\ell$ is large, the mean free path $\ell_{\rm ee}$ for e-e collisions is also very large due to Pauli blocking.
Indeed, normal Fermi liquids at low temperatures~\cite{abrikosovkhalatnikov,Pines_and_Nozieres, Giuliani_and_Vignale} display very large values of $\ell_{\rm ee}$, i.e.~$\ell_{\rm ee} \propto (T_{\rm F}/T)^2$ for temperatures $T \ll T_{\rm F}$, with $T_{\rm F}$ the Fermi temperature.
(In two spatial dimensions there is a very well known~\cite{giuliani_prb_1982} logarithmic correction that has been dropped.)
These severe restrictions, imposed by a rigid Fermi surface on the phase space for two-body e-e collisions, can be relieved by increasing temperature.
Indeed, $\ell_{\rm ee}$ quickly decreases for increasing $T$.
Short e-e mean free paths therefore require to operate at sufficiently elevated temperatures.
At such temperatures, strong scattering of electrons against optical phonons (e.g.~in polar cystals such as GaAs) often leads to the unwanted inequality $\ell < \ell_{\rm ee}$.
Realizing hydrodynamic flow at ``high'' temperatures therefore requires not only ultra-clean crystals but also crystals where electron-phonon coupling is extremely weak. We note that it is in principle easier to reach the hydrodynamic transport regime in non-polar crystals such as graphene, where the dominating mechanism at large temperatures is scattering of electrons against acoustic phonons. In this case $\ell$ decays as $1/T$, i.e.~slower than $\ell_{\rm ee}$.

For these reasons, evidence of hydrodynamic transport in solid-state devices is, to the best of our knowledge, limited to early work carried out by Molenkamp and de Jong~\cite{molenkamp_sse_1994,dejong_prb_1995} in electrostatically defined wires in GaAs/AlGaAs heterostructures.
These authors measured a non-monotonic dependence of the four-point longitudinal resistivity $\rho_{xx}$ on the electronic temperature $T$, which is increased above the lattice temperature by using a large dc heating current.
The decrease of $\rho_{xx}$ with increasing temperature above a certain value of $T$ was attributed to e-e collisions~\cite{molenkamp_sse_1994,dejong_prb_1995}.
This is the so-called ``Gurzhi effect'' and will be discussed extensively in this Article.
More recently, indirect evidence of hydrodynamic flow comes from an explanation~\cite{song_prl_2012} of Coulomb drag between two neutral graphene sheets~\cite{gorbachev_naturephys_2012}, which differs from that offered by the authors of Ref.~\onlinecite{gorbachev_naturephys_2012}.

Recent experimental progress~\cite{mayorov_nanolett_2011,mayorov_nanolett_2012,wang_science_2013,taychatanapat_naturephys_2013,woessner_naturemater_2015,geim_nature_2013}, however, has made it possible to fabricate samples with ultra-high carrier transport mobilities {\it up to room temperature}.
These are graphene sheets encapsulated between thin hexagonal boron nitride (hBN) slabs, which display ultra-high mobilities reaching $10^{5}~{\rm cm}^2/({\rm V} {\rm s})$ in a wide range of temperatures up to room temperature. These values can be achieved in GaAs/AlGaAs heterostructures only below $100~{\rm K}$ because of polar phonon scattering~\cite{pfeiffer_physicaE_2003}. In addition, the finite electron mass and moderate doping required to achieve high mobilities limits the Fermi energy to values below a few tens of meV, which makes the Fermi level smearing important even at liquid nitrogen temperatures.
Encapsulated samples~\cite{gorbachev_naturephys_2012,mayorov_nanolett_2011,mayorov_nanolett_2012,wang_science_2013,taychatanapat_naturephys_2013,woessner_naturemater_2015,geim_nature_2013} are special because electrons roaming in graphene suffer very weak scattering against acoustic phonons~\cite{hwang_prb_2008,principi_prb_2014,park_nanolett_2014,sohier_prb_2014} and because hBN provides an exceptionally clean and flat dielectric environment for graphene~\cite{dean_naturenano_2010}.
Furthermore, microscopic calculations based on many-body diagrammatic perturbation theory~\cite{polini_arxiv_2014,li_prb_2013,principi_arxiv_2015} indicate that the e-e mean free path $\ell_{\rm ee}$ in graphene is shorter than $400~{\rm nm}$ in a wide range of carrier densities and temperatures $T \geq 150~{\rm K}$.
We therefore conclude that hBN/graphene/hBN stacks are ideal samples where the long-sought hydrodynamic regime can be unveiled and explored.
Indeed, recent non-local transport measurements~\cite{bandurin_arxiv_soon} carried out in high-quality encapsulated single-layer (SLG) and bilayer (BLG) graphene samples have demonstrated that this is the case.
The authors of Ref.~\onlinecite{bandurin_arxiv_soon} have reported evidence of hydrodynamic transport, showing that doped graphene exhibits an anomalous (negative) voltage drop near current injection points, which has been attributed to the formation of whirlpools in the electron flow.
From measurements of non-local signals, Bandurin {\it et al.}~\cite{bandurin_arxiv_soon} extracted the viscosity of graphene's electron liquid and found it to be in quantitative agreement with many-body theory calculations~\cite{principi_arxiv_2015}.

In this Article, we present a fully-analytical theoretical study of non-local dc transport in the two-dimensional (2D) electron liquid in a graphene sheet in the hydrodynamic regime. In Sect.~\ref{sect:linearizedtheory} we present the theoretical framework that was used in Ref.~\onlinecite{bandurin_arxiv_soon} to interpret the experimental results, i.e.~a linearized steady-state hydrodynamic approach based on the continuity and Navier-Stokes equations.
Suitable boundary conditions for these hydrodynamic equations are discussed in Sect.~\ref{subsect:BCs}.
In Sect.~\ref{sect:longitudinal} we present analytical solutions for longitudinal transport in a rectangular Hall bar and we discuss the dependence of the solutions on the boundary conditions, providing details on the Gurzhi effect.
In Sect.~\ref{sect:nonlocal} we present analytical results for the spatial depedence of the 2D electrical potential, non-local resistance, and current-induced magnetic field, as obtained by solving the hydrodynamic equations with the free-surface boundary conditions, which we believe to be the appropriate boundary conditions for the linear-response regime.
Finally, in Sect.~\ref{sect:summary} we present a summary of our main results and offer some perspectives.

We remark that, in the present work, we focus only on doped SLG and BLG sheets, where the applicability of the Fermi-liquid theory is granted.
However, it is believed that the hydrodynamic behavior of the semimetals is particularly interesting~\cite{muller_prb_2008,muller_prl_2009} when these are in the charge neutral state.
In this case the Fermi surface shrinks to a point and Fermi liquid theory is not applicable.
For example, the authors of Ref.~\onlinecite{muller_prl_2009} have found that the ratio $\eta/s$ for the 2D electron liquid in a graphene sheet at the charge neutrality point comes close to saturating the AdS/CFT bound.
In nearly neutral semimetals $\ell_{\rm ee}$ is also short due to frequent collisions between thermally excited carriers ($T \gg T_{\rm F}$).
It is, however, very well known~\cite{dassarma_rmp_2011} that any theory at the charge neutrality point must take into account the spatially inhomogeneous pattern of electron-hole puddles created by disorder~\cite{martin_naturephys_2008}.
In the regime of sizeable doping we consider, we can safely ignore puddles in encapsulated graphene devices~\cite{xue_naturemater_2011,decker_nanolett_2011}.

\section{Linearized steady-state hydrodynamic theory}
\label{sect:linearizedtheory}

\begin{figure}[t]
\includegraphics[width=\linewidth]{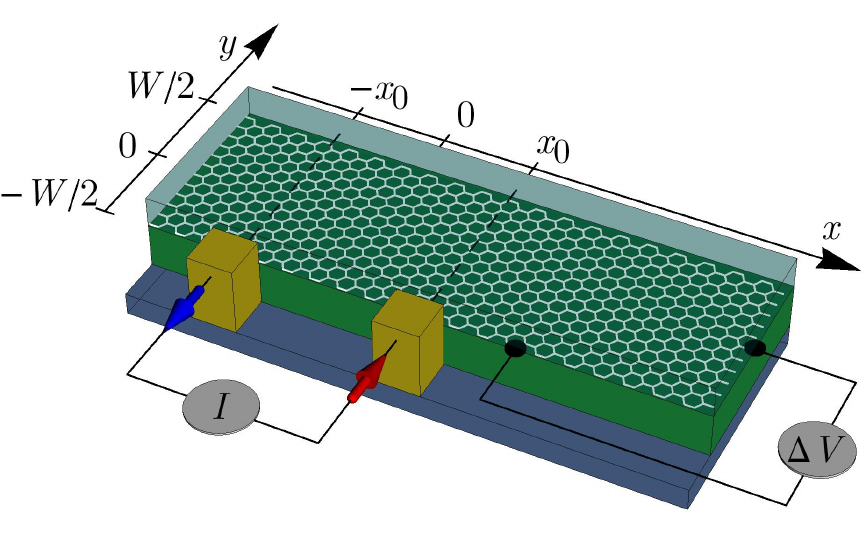}
\caption{\label{fig:Device}
(Color online) Schematic representation of the non-local transport setup analyzed in this work.
A  dc current $I$ is injected (red arrow) into an encapsulated graphene Hall bar of width $W$.
Current injection occurs at a lateral contact located at $x = x_0$ and $y= - W/2$.
The same current is drained (blue arrow) at a contact located at $x = - x_0$ and $y= - W/2$.
Measurements of voltage drops $\Delta V$ ``near'' the current injection region are sensitive to the kinematic viscosity $\nu$ of the two-dimensional massless Dirac fermion liquid.
The notion of  ``vicinity'' between voltage probe and current injector is defined by a crucial length scale, i.e.~the vorticity diffusion length $D_{\nu} = \sqrt{\nu\tau}$.
Here $\tau$ (exceeding $1~{\rm ps}$ in high-quality encapsulated devices) represents a phenomenological scattering time due to momentum-non-conserving collisions of a fluid element (and {\it not} of single electrons). }
\end{figure}

We consider a two-dimensional electron liquid in a doped SLG or BLG sheet, deep in the hydrodynamic transport regime ($\ell_{\rm ee} \ll \ell,W$).
For the sake of definiteness, we consider the Hall bar geometry sketched in Fig.~\ref{fig:Device}.
Since the energy-momentum dispersion of electrons in these systems is particle-hole symmetric~\cite{kotov_rmp_2012}, we assume, without loss of generality, that the sample hosts a back gate-controlled equilibrium electron density equal to $\bar{n}$.
(The charge density is $-e{\bar n}$, $-e$ being the electron charge.)
We neglect thermally-excited carriers and coupling between charge and heat flow~\cite{phan_arxiv_2013}, which is strong only at the charge neutrality point.
Finally, we consider the linear response regime and steady-state transport.

In this framework of approximations, the hydrodynamic transport equations~\cite{tomadin_prb_2013,tomadin_prl_2014} for the 2D electron liquid greatly simplify and reduce to
\begin{equation}\label{eq:ContinuityLinearised}
\nabla \cdot {\bm J}({\bm r})=0~,
\end{equation}
and
\begin{equation}\label{eq:NavierStokesLinearised}
\frac{\bar{n} e}{m}\nabla \phi({\bm r})+ \nu \nabla^2 {\bm J}({\bm r}) = \frac{{\bm J}({\bm r})}{\tau}~.
\end{equation}
In Eqs.~(\ref{eq:ContinuityLinearised}) and~(\ref{eq:NavierStokesLinearised}) we have introduced the linearized steady-state particle current density ${\bm J}({\bm r}) = {\bar n} {\bm v}({\bm r})$, where
${\bm v}({\bm r})$ is the linearized steady-state fluid-element velocity.

Eq.~(\ref{eq:ContinuityLinearised}) is the continuity equation, while Eq.~(\ref{eq:NavierStokesLinearised}) is the Navier-Stokes equation.
The latter contains three forces acting on a fluid element: i) the electric force $-e{\bm E}({\bm r}) = e \nabla \phi({\bm r})$, written in terms of the electric potential $\phi({\bm r})$ in the 2D plane where electrons are moving, which is generated by the steady-state charge distribution $n({\bm r})$ in response to the drive current $I$; ii) the internal force due to the shear viscosity $\eta = \eta(\bar{n},T)$ of the 2D electron liquid, here written in terms of the kinematic viscosity~\cite{landaufluidmechanics,batchelor,huangbook}
\begin{equation}\label{eq:NuDefinition}
\nu = \frac{\eta}{m \bar{n}}~;
\end{equation}
and iii) friction exerted on a fluid element by agents external to the electron liquid such as phonons and impurities, which dissipate the fluid-element momentum at a rate $\tau^{-1} = 1/\tau(\bar{n},T)$.
The latter is a phenomenological parameter, which depends on $\bar{n}$ and $T$ and is commonly used in modelling transport in semiconductor devices~\cite{engineering_books}.

In Eqs.~(\ref{eq:NavierStokesLinearised}) and~(\ref{eq:NuDefinition}) $m$ is a suitable effective mass defined by:
\begin{equation}\label{eq:mass}
m =
\left\{\begin{array}{l}
m_{\rm c},~{\rm for~SLG},\vspace{0.2 cm}\\
0.03~m_{\rm e},~{\rm for~BLG}
\end{array}
\right.~,
\end{equation}
where $m_{\rm c} = \hbar k_{\rm F}/v_{\rm F}$ is the 2D massless Dirac fermion cyclotron mass~\cite{kotov_rmp_2012}, $k_{\rm F} = \sqrt{\pi \bar{n}}$ being the Fermi wave number and $v_{\rm F}\sim 10^6~{\rm m}/{\rm s}$ the Fermi velocity, and $m_{\rm e}$ is the bare electron mass in vacuum.

Multiplying both members of Eq.~(\ref{eq:NavierStokesLinearised}) by $\tau$, we obtain
\begin{equation}\label{eq:NavierStokesLinearised_2}
\frac{\sigma_0}{e}\nabla \phi({\bm r})+ D^2_{\nu}\nabla^2 {\bm J}({\bm r}) = {\bm J}({\bm r})~.
\end{equation}
In Eq.~(\ref{eq:NavierStokesLinearised_2}) we have introduced the following characteristic length scale of the problem:
\begin{equation}\label{eq:vorticity_diffusion_constant}
D_{\nu} \equiv \sqrt{\nu\tau}~.
\end{equation}
For $\tau = 1~{\rm ps}$ (as in high-quality hBN/graphene/hBN samples) and $\nu = 0.1~{\rm m}^2/{\rm s}$ (see Ref.~\onlinecite{principi_arxiv_2015}) we obtain $D_{\nu} \approx 0.3~{\rm \mu m}$.

The physical significance of $D_{\nu}$ can be understood as follows.
We first note that we can rewrite $\nabla^2 {\bm J}({\bm r})$ by using the following identity:
\begin{equation}\label{eq:identity}
\nabla^2 {\bm J}({\bm r}) = \nabla [\nabla\cdot {\bm J}({\bm r})] - \nabla \times [\nabla \times {\bm J}({\bm r})]~.
\end{equation}
Because of (\ref{eq:ContinuityLinearised}), we can drop the first term on the right-hand side of Eq.~(\ref{eq:identity}). The second term is finite and related to the vorticity~\cite{landaufluidmechanics,batchelor}
\begin{equation}\label{eq:vorticity}
\bm{\omega}({\bm r}) \equiv \frac{1}{\bar{n}}\nabla \times {\bm J}({\bm r}) = \omega({\bm r}) \hat{\bm z}~,
\end{equation}
which in 2D is oriented along the $\hat{\bm z}$ axis. We can then rewrite Eq.~(\ref{eq:NavierStokesLinearised_2}) as following
\begin{equation}\label{eq:intermediate}
\frac{\sigma_0}{e}\nabla \phi({\bm r}) - {\bar n} D^2_{\nu}\nabla \times \bm{\omega}({\bm r}) = {\bm J}({\bm r})~.
\end{equation}
Taking the curl of Eq.~(\ref{eq:intermediate}) and using the identities $\nabla \times \nabla \phi({\bm r}) = 0$ and $\nabla \cdot {\bm \omega}({\bm r}) = 0$ (the latter being valid because ${\bm \omega}$'s only non-vanishing component is along $\hat{\bm z}$, while $\nabla$ acts only on the 2D $\hat{\bm x}$-$\hat{\bm y}$ plane), we obtain a damped-diffusion equation for the vorticity
\begin{equation}\label{eq:diffusionvorticity}
D_{\nu}^2 \nabla^2 \omega(\bm{r})=\omega(\bm{r})~.
\end{equation}
We therefore see that $D_{\nu}$ plays the role of a diffusion length for $\omega(\bm{r})$.

In Eq.~(\ref{eq:NavierStokesLinearised_2}) we have also introduced a ``Drude-like'' conductivity,
\begin{equation}
\sigma_0 \equiv \frac{e^2 \bar{n}\tau}{m}~.
\end{equation}
Since we are in the hydrodynamic regime, $\sigma_0$ should not be confused~\cite{tomadin_prl_2014} with the ordinary dc conductivity in the diffusive transport regime: once again, $\tau = \tau(\bar{n},T)$ represents a phenomenological parameter that should be fit to experimental data, as we discuss below in Sect.~\ref{sect:longitudinal}.
This na\"{i}ve description of momentum-non-conserving collisions in the hydrodynamic transport regime can be relaxed by following similar arguments to those in Ref.~\onlinecite{andreev_prl_2011}: this is however well beyond the scope of the present Article and will be the topic of future studies.
In the absence of viscosity, Eq.~(\ref{eq:NavierStokesLinearised}) reduces to a local version of Ohm's law, i.e.~${\bm J}({\bm r}) = \sigma_0 \nabla \phi({\bm r})/e$.

Finally, we note that taking the divergence of Eq.~(\ref{eq:NavierStokesLinearised_2}) and making use of Eq.~(\ref{eq:ContinuityLinearised}) we obtain the Laplace equation $\nabla^2 \phi({\bm r}) =0$ for the electric potential $\phi({\bm r})$ on the 2D plane.
This should not be confused with the usual three-dimensional (3D) Poisson equation for the 3D electrostatic potential $\Phi({\bm r}, z)$,
\begin{equation}\label{eq:poisson}
\left(\nabla^2+ \frac{\partial^2}{\partial z^2}\right)\Phi({\bm r}, z) = 4\pi e n({\bm r})\delta(z)~.
\end{equation}
The 2D potential in Eq.~(\ref{eq:NavierStokesLinearised}) is $\phi({\bm r}) = \Phi({\bm r}, z=0)$.
On the right-hand side of Eq.~(\ref{eq:poisson}) we note the steady-state charge density distribution $-en({\bm r})$ which occurs in the sample in response to the drive current $I$.
Eq.~(\ref{eq:poisson}) needs to be solved in 3D space with suitable boundary conditions---depending on the dielectric environment, gates, etc.~surrounding the graphene sheet---if one is interested in determining $n({\bm r})$.
In this Article we will focus our attention on ${\bm J}({\bm r})$ and $\phi({\bm r})$.

Eqs.~(\ref{eq:ContinuityLinearised}) and~(\ref{eq:NavierStokesLinearised_2}) will be used to describe transport in the Hall bar geometry pictorially represented in Fig.~\ref{fig:Device}.
Mathematically, it is convenient to work in a Hall bar of infinite length in the longitudinal direction $\hat{\bm x}$, since this allows us to use the Fourier transform to solve the equations of motion---see Sect.~\ref{sect:nonlocal}.
The width $W$ of the Hall bar will be kept finite.
In the next Section we will describe a crucially important ingredient of the theory: boundary conditions.

\subsection{Boundary conditions}
\label{subsect:BCs}

In order to find $\phi({\bm r})$ and ${\bm J}({\bm r})$ in the Hall bar geometry depicted in Fig.~\ref{fig:Device}, we need to solve Eqs.~(\ref{eq:ContinuityLinearised}) and~(\ref{eq:NavierStokesLinearised_2}) in the rectangle $(-\infty,\infty) \times [-W/2,W/2]$, with appropriate boundary conditions (BCs) at the edges, i.e.~at $y=\pm W/2$.

Lateral electrodes acting as current injectors/collectors are described through BCs on the component of the current {\it perpendicular} to the edges:
\begin{equation}\label{eq:NormalBC}
J_y\left(x,y=\pm W/2\right)={\cal J}_{\pm}(x)~.
\end{equation}
Here ${\cal J}_{\pm}(x)$ is a function that describes a distribution of current injectors and collectors on the upper (lower) edge of the multi-terminal Hall bar. It is through Eq.~(\ref{eq:NormalBC}) that the total drive current $I$ injected into the system at the boundaries enters the problem.

Following Abanin {\it et al.}~\cite{abanin_science_2011}, we model the electrodes as point-like (i.e.~delta-function) sources and sinks.
(A more realistic modeling of electrodes has been carried out in Ref.~\onlinecite{bandurin_arxiv_soon}, where finite-width effects and metallic boundary conditions at extended electrodes have been taken into account in a fully numerical solution of Eqs.~(\ref{eq:ContinuityLinearised}) and~(\ref{eq:NavierStokesLinearised_2}).
Such details have essentially no impact on the physics we are going to highlight below.)
For example, for the setup depicted in Fig.~\ref{fig:Device} with a current injector at $x=x_0$, a current collector at $x = -x_0$, and no injectors/collectors on the upper edge, we will use:
\begin{equation}\label{eq:BCpointlike}
{\cal J}_{-}(x) = - \frac{I}{e}\delta(x-x_0) + \frac{I}{e}\delta(x+x_0)~,
\end{equation}
and ${\cal J}_{+}(x)=0$.

In the presence of a finite shear viscosity $\nu$, we need an additional BC on the {\it tangential} component of the current at the top ($y = +W/2$) and bottom ($y = -W/2$) edges of the Hall bar.
We use the following BC:
\begin{equation}\label{eq:generalBC}
\left[\partial_y J_x(x,y) + \partial_x J_y(x,y)\right]_{y = \pm W/2}= \mp \frac{J_x(x,y = \pm W/2)}{l_{\rm b}}~,
\end{equation}
where $l_{\rm b}$ is a ``boundary slip length'', i.e.~a length scale describing friction at the physical boundaries of the sample.
This BC can be explained as following.
The left-hand side of Eq.~(\ref{eq:generalBC}) is proportional to the off-diagonal component of the stress tensor~\cite{landaufluidmechanics}, calculated at the edges of the Hall bar.
It represents the tangential component of the frictional force exerted by the boundaries of the Hall bar on the 2D electron liquid~\cite{landaufluidmechanics}.
This force depends on the tangential velocity of the 2D electron liquid and boundary roughness: in the linear-response regime, it is natural to replace such unknown dependence with a linear law characterized by the single parameter $l_{\rm b}$, as in the right-hand side of Eq.~(\ref{eq:generalBC}).

In the description of transport of molecular liquids in constrained geometries, like water in a pipe, where the interactions between the molecules of the fluid and the walls of the container are of the same nature as of those between molecules of the fluid, the most used BCs are the so-called ``no-slip'' BCs~\cite{landaufluidmechanics}, in which the component of the current tangential to the boundary vanishes.
The no-slip BCs can be obtained from Eq.~(\ref{eq:generalBC}) by taking the limit $l_{\rm b} \to 0$. 
In the opposite limit of a free-surface geometry, like the surface of water in an open bucket, the tangential force applied from the boundary to the fluid element vanishes at the boundary. These ``free-surface'' BCs~\cite{landaufluidmechanics,tomadin_prl_2014} can be obtained from Eq.~(\ref{eq:generalBC}) by taking the limit $l_{\rm b} \to +\infty$.

Which of these BCs should be used to model the experiments in Ref.~\onlinecite{bandurin_arxiv_soon} will become clear at the end of Sect.~\ref{sect:longitudinal}.

\subsection{Applicability of the linearized theory}
\label{sect:smallness}

The validity of the linearized Navier-Stokes equation (\ref{eq:NavierStokesLinearised_2}) relies on the smallness of the Reynolds number~\cite{landaufluidmechanics,batchelor,huangbook} ${\cal R}_W$.
This is a dimensionless parameter (which depends on the sample geometry) that controls the smallness of the non-linear term $[{\bm v}({\bm r},t) \cdot \nabla]{\bm v}({\bm r},t)$ in the convective derivative with respect to the viscous term.
In our case we can define the Reynolds number as following:
\begin{equation}\label{eq:reynoldsnumber}
\left|\frac{[{\bm v}({\bm r},t) \cdot \nabla] {\bm v}({\bm r},t)}{\nu \nabla^2 {\bm v}({\bm r},t)}\right| \simeq \frac{\bar{v}W}{\nu} = \frac{I}{e {\bar n} \nu} \equiv {\cal R}_W~,
\end{equation}
where ${\bar v}$ is the typical value of the fluid-element velocity.
For an injected current~\cite{bandurin_arxiv_soon} $I=2 \times 10^{-7}~{\rm A}$, a Hall bar width $W=1~{\rm \mu m}$, and an equilibrium density $\bar{n}=10^{12}~{\rm cm}^{-2}$, we obtain $\bar{v} \sim I/(e \bar{n} W) \approx 10^{4}~{\rm cm}/{\rm s}$.
We note that $\bar{v}$ is much smaller than the graphene Fermi velocity $v_{\rm F} \sim 10^6~{\rm m}/{\rm s}$ and the flow is therefore ``non-relativistic.''
The corresponding value of the Reynolds number is ${\cal R}_W \sim 10^{-3} \ll 1$, obtained by using a kinematic viscosity $\nu \sim 10^3~{\rm cm^2/s}$ of the 2D electron liquid in graphene~\cite{principi_arxiv_2015}.
Our linearized theory in Eqs.~(\ref{eq:ContinuityLinearised}) and~(\ref{eq:NavierStokesLinearised_2}) is therefore fully justified.

\section{Longitudinal transport and the Gurzhi effect}
\label{sect:longitudinal}

We first consider the situation in which no current is injected or extracted laterally at the Hall bar edges, i.e.~${\cal J}_{\pm}(x)=0$.

In this case the local current ${\bm J}({\bm r})$ does not depend on the longitudinal coordinate $x$ and all the spatial derivatives with respect to $x$ in Eqs.~(\ref{eq:ContinuityLinearised}), (\ref{eq:NavierStokesLinearised_2}), and (\ref{eq:generalBC}) vanish.
The continuity equation implies that $J_y$ does not depend on $y$ and vanishes identically because of Eq.~(\ref{eq:NormalBC}).
Therefore also the $y$ component of the electric field must vanish.
The $x$ component of the current respects the following equation: $J_x(y)-D^2_\nu \partial_y^2 J_x(y)= - \sigma_0 E_x/e$,  where ${\bm E} = - \nabla \phi({\bm r})$ is the electric field.
Note that $E_x$ cannot depend on $y$ because $E_y$ vanishes and $\nabla \times {\bm E}=0$.
The solution of this equation that fulfils the BC (\ref{eq:generalBC}) is
\begin{equation}\label{eq:SolutionUniform}
J_x(y)=-\frac{\sigma_0}{e}E_x\left[ 1- \frac{D_{\nu}}{\xi}\cosh{\left (\frac{y}{D_{\nu}} \right )} \right]~,
\end{equation}
where we have introduced the length
\begin{equation}\label{eq:denominator_form_factor}
\xi \equiv l_{\rm b} \sinh\left(\frac{W}{2 D_{\nu}}\right)+ D_{\nu}\cosh\left(\frac{W}{2D_{\nu}}\right)~.
\end{equation}
We can calculate the total longitudinal current $I$ carried by the flow by integrating Eq.~(\ref{eq:SolutionUniform}) in the transverse direction, i.e.
\begin{equation}\label{eq:TotalCurrent}
I =-e\int_{-W/2}^{W/2}dyJ_x(y) = \sigma_{0} W E_x (1 - {\cal F})~,
\end{equation}
where we have defined the dimensionless quantity
\begin{equation}\label{eq:form-factor}
{\cal F} \equiv 2\frac{D^2_{\nu}}{W \xi} \sinh\left(\frac{W}{2D_{\nu}}\right)~.
\end{equation}
Measuring the longitudinal potential drop $\Delta V$ between two lateral contacts at positions $x$ and $x+L$ yields a four-point longitudinal conductivity $\sigma_{xx}$ of the form:
\begin{equation}\label{eq:ViscousConductanceCorrections}
\sigma_{xx} \equiv \frac{I}{\Delta V}\frac{L}{W} = \sigma_0(1-{\cal F})~.
\end{equation}
Eq.~(\ref{eq:ViscousConductanceCorrections}) is the most important result of this Section.

\begin{figure}[t]
\begin{overpic}[width=\linewidth]{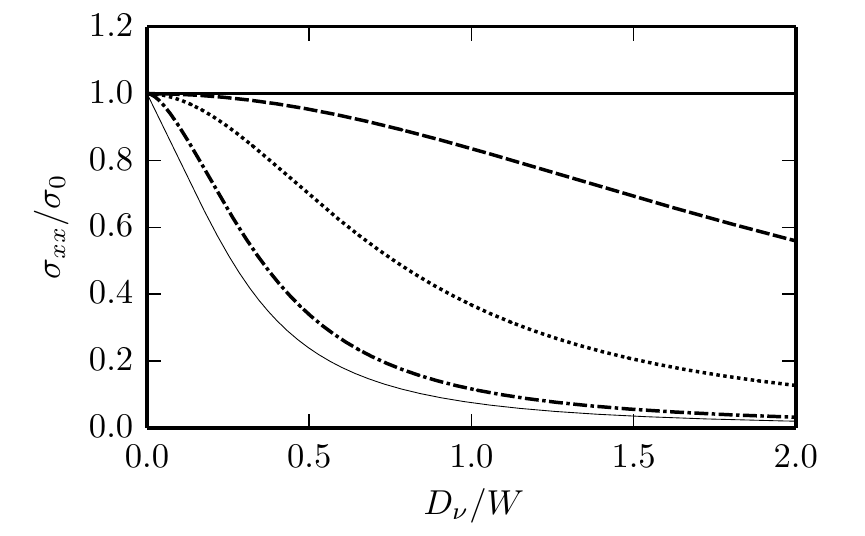}\put(2,140){(a)}\end{overpic}
\begin{overpic}[width=\linewidth]{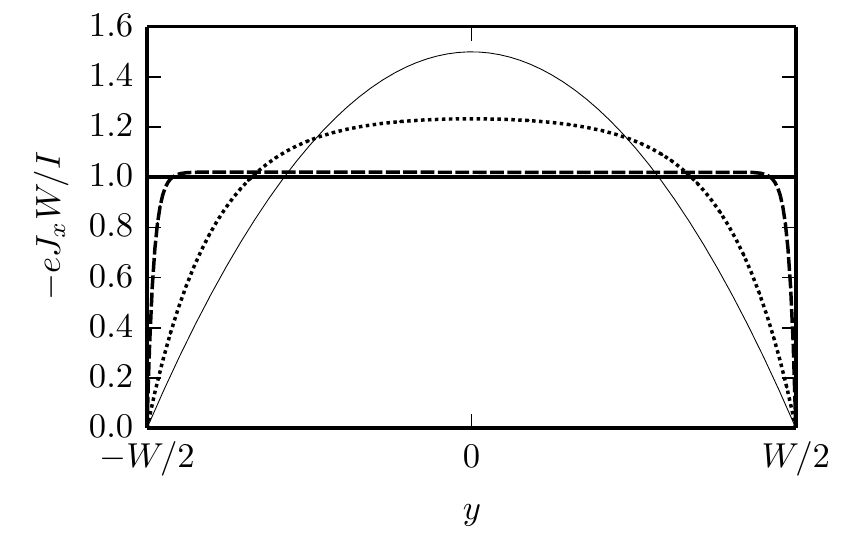}\put(2,140){(b)}\end{overpic}
\caption{\label{fig:longitudinal}
Panel (a) The longitudinal conductivity (\ref{eq:ViscousConductanceCorrections}) (in units of $\sigma_0$) is plotted as a function of the ratio $D_{\nu}/W$ for different values of the boundary scattering length $l_{\rm b}$: $l_{\rm b} = \infty$ (thick solid line), $l_{\rm b} = 10~W$ (dashed line), $l_{\rm b} = W$ (dotted line), $l_{\rm b} = 0.1~W$ (dash-dotted line), and $l_{\rm b} = 0$ (thin solid line).
Panel (b) The current-density profile $-eJ_x(y)$, normalized by the total current $I/W$, is plotted as a function of $y$ for $l_{\rm b}=0$ (no-slip BCs) and different values of $D_{\nu}$: $D_{\nu} = 0$ (thick solid line), $D_{\nu} = 0.01~W$ (dashed line), and $D_{\nu} = 0.1~W$ (dotted line).
The case $D_{\nu}\gg W$, corresponding to Poiseuille flow, is represented by a thin solid line. }
\end{figure}

In the limit $l_{\rm b} \to \infty$ (i.e.~free-surface BCs) ${\cal F} \to 0$.
For this choice of BCs the longitudinal conductivity $\sigma_{xx}$ depends only on the rate of momentum-non-conserving collisions $\tau^{-1}$ (through $\sigma_0$) and is independent of $\nu$.

On the other hand, in the limit $l_{\rm b} \to 0$ (i.e.~no-slip BCs) Eq.~(\ref{eq:ViscousConductanceCorrections}) reduces to
\begin{equation}\label{eq:SigmaNoSlip}
\sigma_{xx} = \sigma_0\left[1-2\frac{D_{\nu}}{W}\tanh\left(\frac{W}{2D_{\nu}}\right) \right]~.
\end{equation}
We can easily understand two asymptotic limits of Eq.~(\ref{eq:SigmaNoSlip}).
In the limit $D_{\nu}\ll W$ Eq.~(\ref{eq:SigmaNoSlip}) yields $\sigma_{xx} = \sigma_0(1-2D_{\nu}/W)$: the small correction to the Drude-like conductivity $\sigma_0$ is due to a reduction of the fluid-element velocity in a thin region of width $D_{\nu}$ near the top and bottom edges of the Hall bar.
In the opposite limit, $D_{\nu} \gg W$, we obtain $\sigma_{xx}= \sigma_0 W^2/(12 D^2_{\nu}) = e^2 \bar{n}W^2/(12 m \nu)$.
In this limit the problem is equivalent to that of Poiseuille flow in a pipe~\cite{landaufluidmechanics,batchelor}, with a velocity profile $v_x(y)$ that depends quadratically on the transverse coordinate $y$ and a resistance that is entirely due to viscosity.

A summary of our main results for longitudinal electron transport in the presence of a finite viscosity is reported in Fig.~\ref{fig:longitudinal}.

\subsection{The Gurzhi effect}

We now would like to make a remark on the temperature dependence of $\sigma_{xx}$ in Eq.~(\ref{eq:ViscousConductanceCorrections}).
For the sake of simplicity, we assume that $l_{\rm b}$ does not depend on temperature.
We observe that the derivative of $\sigma_{xx}$ with respect to $T$,
\begin{equation}\label{eq:T_dependence}
\frac{d \sigma_{xx}}{dT} = \frac{d\sigma_0}{dT}(1- {\cal F}) - \sigma_0 \frac{d {\cal F}}{d D_{\nu}} \frac{d D_{\nu}}{dT}~,
\end{equation}
is the sum of two contributions with opposite signs.
The first term on the right-hand side of Eq.~(\ref{eq:T_dependence}) is negative, because ${\cal F} < 1$ and $d\sigma_0 / dT < 0$. 
The latter inequality holds because the scattering rate $\tau^{-1}$ describing momentum non-conserving collisions is a monotonically increasing function of temperature~\cite{bandurin_arxiv_soon}.
On the contrary, the second term is positive, because the $d {\cal F} / d D_{\nu} > 0$ and $dD_{\nu} / dT<0$.
The vorticity diffusion length $D_{\nu}$ decreases with increasing temperature because both $\nu$ and $\tau$ are decreasing functions of $T$.
We therefore conclude that, due to viscosity, $\sigma_{xx}$ ($\rho_{xx}$) can increase (decrease) upon increasing temperature.
This is the so-called Gurzhi effect~\cite{gurzhi_spu_1968}.
The existence of this effect relies crucially on the nature of BCs that are used to solve the hydrodynamic equations.
In particular, it disappears for free-surface BCs. All previous experimental studies of transport in graphene and other 2D electron liquids we are aware of have reported monotonic temperature dependencies (i.e.~no evidence of the Gurzhi effect) in the ordinary longitudinal geometry in the linear-response regime. We therefore conclude that free-surface BCs are the most appropriate for a weak driving current $I$.
In this case, $\sigma_{xx}$ depends only on the unknown damping rate $\tau^{-1}$, which can therefore be determined from an ordinary four-point longitudinal transport measurement at every value of $\bar{n}$ and $T$, i.e.~$\tau^{-1}=  e^2 {\bar n}/(m \sigma_{xx})$.

In the next Section, we will discuss another hydrodynamic phenomenon occurring in 2D electron liquids, i.e.~the formation of whirlpools in electron flow~\cite{bandurin_arxiv_soon}, yielding a clear-cut experimental signal of hydrodynamic transport in weakly non-local linear-response transport measurements.
Since the experimental data in Ref.~\onlinecite{bandurin_arxiv_soon} do not show any Gurzhi effect in the linear-response regime, in the next Section we will utilize only the free-surface BCs ($l_{\rm b} \to \infty$).
Whirlpools in hydrodynamic electron flow, however, do exist also when no-slip BCs are used~\cite{bandurin_arxiv_soon}. In this sense whirpools are a much more robust phenomenon that the Gurzhi effect in longitudinal transport. Whirpools are also more dramatic in experimental appearance.

\section{Non-local transport and the impact of viscosity}
\label{sect:nonlocal}

We now present the solution of the problem posed by Eqs.~(\ref{eq:ContinuityLinearised}), (\ref{eq:NavierStokesLinearised_2}), (\ref{eq:NormalBC}), and (\ref{eq:generalBC}) in the rectangle $(-\infty,\infty) \times [-W/2,W/2]$. We use free-surface boundary conditions, corresponding to $l_{\rm b}\to \infty$.

To this end, we introduce the Fourier transform with respect to the longitudinal coordinate $x$ in the three equations~(\ref{eq:ContinuityLinearised}) and~(\ref{eq:NavierStokesLinearised_2}) (the latter for the two components $J_x$ and $J_y$). The Fourier transform of a function $f(x,y)$ will be denoted by $\ft{f}(k,y)$.
These equations can be grouped into a linear system of three second-order ordinary differential equations (ODEs) with respect to the independent variable $y$.
It is convenient to rewrite this system in terms of four first-order ODEs.
We find
\begin{widetext}
\begin{equation}\label{eq:System}
\partial_{y}
\left ( \begin{array}{c}
k \ft{J}_x(k,y) \\
k \ft{J}_y(k,y) \\
\partial_y \ft{J}_x(k,y) \\
k^{2} \sigma_{0} \ft{\phi}(k,y) / e \\
\end{array} \right )
=
k \left ( \begin{array}{cccc}
0 &0 & 1 &0 \\
-i &0 &0 &0 \\
1 + 1/(k D_{\nu})^{2} & 0 & 0 & -i / (k D_{\nu})^{2} \\
0 & 1 + (k D_{\nu})^{2} & i (k D_{\nu})^{2} & 0 \\
\end{array} \right )
\left ( \begin{array}{c}
k \ft{J}_x(k,y) \\
k \ft{J}_y(k,y) \\
\partial_y \ft{J}_x(k,y) \\
k^{2} \sigma_0 \ft{\phi}(k,y) / e \\
\end{array} \right )~.
\end{equation}
\end{widetext}
Eq.~(\ref{eq:System}) can be solved by diagonalizing the $4\times4$ matrix on the right-hand side, which has four distinct eigenvalues: $\pm 1$ and $\pm \sqrt{1+1/(k D_{\nu})^{2}}$.
The general solution will therefore be a linear combination of exponentials of the form $\sum_i a_i v_i \exp(\lambda_i k y)$ where $v_i$ ($\lambda_i$) are the eigenvectors (eigenvalues) of the matrix.
The four unknown coefficients $a_i$ can be found by enforcing the desidered BCs.
These are found by taking the Fourier transform of Eqs.~(\ref{eq:NormalBC}) and~(\ref{eq:generalBC}) with respect to $x$:
\begin{equation}\label{NormalBCFourier}
\ft{J}_y(k, y = \pm W/2)=\ft{\cal{J}}_{\pm}(k)
\end{equation}
and 
\begin{equation}\label{NoSlipBCFourier}
[\partial_y \ft{J}_x(k,y) + ik \ft{J}_y(k,y)]_{y = \pm W/2}=0~.
\end{equation}
The solution reads as follows:
\begin{widetext}
\begin{equation}\label{eq:potential_infinite_strip}
\ft{\phi}(k,y)=\sum_{\alpha=\pm}  \frac{e\ft{\cal J}_{\alpha}(k)W}{\sigma_0}\left[\ft{F}_{1\alpha}\left(kW,\frac{y}{W}\right)+\frac{2D_{\nu}^2}{W^2} \ft{F}_{2\alpha}\left(kW,\frac{y}{W}\right)\right]~,
\end{equation}
\begin{equation}\label{eq:current_x_infinite_strip}
\ft{J_x}(k,y) =\sum_{\alpha=\pm}\ft{\cal J}_{\alpha}(k)W\times \left
\{ik \left[\ft{F}_{1\alpha}\left(kW,\frac{y}{W}\right)+\frac{2D_{\nu}^2}{W^2} \ft{F}_{2\alpha}\left(kW,\frac{y}{W}\right)\right]-\frac{2D_{\nu}^2}{W^2} \partial_y \ft{F}_{3\alpha}\left(kW,\frac{y}{W},\frac{D_{\nu}}{W}\right)\right\}~,
\end{equation}
and
\begin{equation}\label{eq:current_y_infinite_strip}
\ft{J_y}(k,y) =\sum_{\alpha=\pm} \ft{\cal J}_{\alpha}(k)W\times
\left\{\partial_y \left[\ft{F}_{1\alpha}\left(kW,\frac{y}{W}\right)+\frac{2D_{\nu}^2}{W^2} \ft{F}_{2\alpha}\left(kW,\frac{y}{W}\right)\right]+\frac{2D_{\nu}^2}{W^2} ik \ft{F}_{3\alpha}\left(kW,\frac{y}{W},\frac{D_{\nu}}{W}\right)\right\}~.
\end{equation}
\end{widetext}
In writing Eqs.~(\ref{eq:potential_infinite_strip})-(\ref{eq:current_y_infinite_strip}) we have introduced the following functions of dimensionless arguments:
\begin{equation}\label{eq:F1}
\ft{F}_{1\pm}(\ks,\ys) =\frac{1}{2} \left[\frac{\sinh(\ks \ys)}{\ks \cosh(\ks/2)} \pm \frac{\cosh(\ks\ys)}{\ks \sinh(\ks/2)}\right]~,
\end{equation}
\begin{equation}\label{eq:F2}
\ft{F}_{2\pm}(\ks,\ys)= \frac{\ks}{2}\left[\frac{\sinh(\ks\ys)}{\cosh(\ks/2)} \pm  \frac{\cosh(\ks\ys)}{\sinh(\ks/2)}\right]~,
\end{equation}
and
\begin{eqnarray}\label{eq:F3}
\ft{F}_{3\pm}(\ks,\ys,\lambda) &=& \frac{i\ks}{2} \left[\frac{\cosh(\ys\sqrt{\ks^2+\lambda^{-2}})}{\cosh(1/2\sqrt{\ks^2+\lambda^{-2}})}\right. \nonumber\\
& \pm & \left.\frac{\sinh(\ys\sqrt{\ks^2+\lambda^{-2}})}{\sinh(1/2\sqrt{\ks^2+\lambda^{-2}})} \right]~.
\end{eqnarray}
Eqs.~(\ref{eq:potential_infinite_strip})-(\ref{eq:F3}) are the most important results of this Article.

In general, it is not an easy task to inverse Fourier transform Eqs.~(\ref{eq:potential_infinite_strip})-(\ref{eq:current_y_infinite_strip}) to real space, after the functions $\ft{J}_{\pm}(k)$ have been specified. Indeed, this requires to calculate a convolution which involves the BCs and the functions $F_{1\pm}(\xs,\ys)$, $F_{3\pm}(\xs,\ys)$, and $F_{3\pm}(\xs,\ys,\lambda)$ in real space. We now introduce the inverse Fourier transforms of the functions $\ft{F}_{1\pm}(\ks,\ys)$, $\ft{F}_{2\pm}(\ks,\ys)$, and $\ft{F}_{3\pm}(\ks,\ys,\lambda)$, which read:
\begin{widetext}
\begin{equation}\label{eq:fhexpressions_1}
F_{1\pm}(\xs,\ys)= \frac{1}{4\pi}\ln\left[\frac{1+e^{-2\pi |\xs|}+2\sin(\pi \ys)e^{-\pi |\xs|}}{1+e^{-2\pi |\xs|}-2\sin(\pi \ys)e^{-\pi |\xs|}}\right] \mp \frac{1}{4 \pi} \ln \left[ 1 + e^{-4 \pi |\xs|}+ 2 \cos( 2 \pi \ys) e^{-2 \pi |\xs|} \right] \mp \frac{|\xs|}{2}~,
\end{equation}
and
\begin{eqnarray}\label{eq:fhexpressions_2}
F_{2\pm}(\xs,\ys) &=& \Bigg\{- \pi \sin(\pi \ys)e^{-\pi |\xs|}(1+e^{-2\pi |\xs|})
\left\{1+e^{-4\pi |\xs|}-2[\cos(2 \pi \ys)+2]e^{-2\pi |\xs|}\right\}  \nonumber\\
& & \pm 2 \pi e^{-2 \pi |\xs|}\left[\cos(2 \pi \ys)(1 + e^{-4 \pi |\xs|})+2 e^{-2 \pi |\xs|}\right]\Bigg\} \times
\left[1+e^{-4 \pi |\xs|}+2\cos(2 \pi \ys) e^{-2 \pi |\xs|}\right]^{-2}~.
\end{eqnarray}
The functions $F_{3\pm}(\xs,\ys,\lambda)$ do not have simple expressions in terms of elementary functions but can be cast in the form of exponentially converging series:
\begin{eqnarray}\label{eq:f3h3Definitions}
F_{3\pm}(\xs,\ys,\lambda) &=& - \pi~{\rm sgn}(\xs)
\Bigg\{ \sum_{\ell=0}^{\infty} (2\ell+1)(-1)^{\ell} \cos[(2\ell+1) \pi \ys] e^{-|\xs|\sqrt{\lambda^{-2}+\pi^2 (2\ell+1)^2}}\nonumber\\
&\mp & \sum_{\ell=1}^{\infty} (2\ell)(-1)^{\ell} \sin(2\ell \pi \ys) e^{-|\xs|\sqrt{\lambda^{-2}+\pi^2 (2\ell)^2}}  \Bigg\}~.
\end{eqnarray}
\begin{table}[t]
\begin{tabular}{|c|c|c|c|c|}\hline
                    &                                                 & \multicolumn{2}{c|}{$|\xs|\ll 1$}                       & $|\xs| \gg 1$ \\
                    & $\ys=\mp 1/2 $                                  & $\ys=-1/2$                       & $\ys=1/2$            & $\ys=\mp 1/2$ \\\hline
$F_{1-}$   & $\pi^{-1}\ln(1\mp e^{-\pi |\xs|}) + |\xs|/2$    & $ \pi^{-1}\ln(\pi |\xs|)$        & $\pi^{-1} \ln(2)$    & $|\xs|/2 \mp \pi^{-1} e^{- \pi |\xs|}$ \\\hline
$F_{2-}$   & $\pm \pi e^{-\pi|\xs|}/(1\mp e^{-\pi|\xs|})^2$  & $(\pi \xs^2)^{-1}$               & $-\pi/4$             & $\pm \pi e^{- \pi |\xs|}$ \\\hline
$F_{3-}$   & $(1\pm 1)\delta'(\xs)/2$                        & $\delta'(\xs)$                   & $0$                  & $0$ \\\hline
\end{tabular}
\caption{\label{tab:AsymptoticBehaviours} Explicit expressions of the functions $F_{m-}$ defined in the main text, evaluated at $\ys=\mp 1/2$.
We also summarize useful asymptotic behaviors in the limit $|\xs| \ll 1$ and $|\xs| \gg 1$.
Similar expressions can also be obtained for the quantities $F_{m+}(\xs,\ys)$ by noting that $F_{m+}(\xs,\ys)=F_{m-}(\xs,-\ys)$. }
\end{table}
\end{widetext}
In the following we will make use of a number of asymptotic behaviors of the functions $F_{1\pm}(\xs,\ys)$, $F_{2\pm}(\xs,\ys)$, and $F_{3\pm}(\xs,\ys, \lambda)$, which have been listed for the sake of convenience in Table~\ref{tab:AsymptoticBehaviours}.

The task of calculating the potential and currents in real space simplifies substantially if the currents ${\cal J}_{\pm}(x)$ in Eq.~(\ref{eq:NormalBC}) can be represented by the sum of a finite number of delta-functions in real space. Let us focus on the geometry of Fig.~\ref{fig:Device}, where the Fourier transform of the BCs~(\ref{eq:BCpointlike}) reads $\ft{\cal J}_{-}(k) = - I (e^{i k x_{0}} - e^{-i k x_{0}}) / e$ and $\ft{\cal J}_{+}(k) = 0$. In this case, we find that the steady-state current pattern can be written as
\begin{equation}\label{eq:SolutionDecomposition}
\bm{J}({\bm r})=\frac{\sigma_0}{e}\nabla \phi({\bm r})-\bar{n}D_{\nu}^2\nabla \times {\bm \omega}({\bm r})~,
\end{equation}
where the potential $\phi({\bm r})$ and vorticity ${\bm \omega}({\bm r})\equiv \hat{\bm z}\omega({\bm r})$ are given by:
\begin{eqnarray}\label{eq:SolutionsDeviceGeometryPotential}
\phi({\bm r}) &=&-\frac{I}{\sigma_0}
\Bigg\{ F_{1-}\left(\frac{x_{-}}{W},\frac{y}{W}\right)- F_{1-}\left(\frac{x_{+}}{W},\frac{y}{W}\right)\nonumber\\
&+&\frac{2D_{\nu}^2}{W^2} \left[F_{2-}\left(\frac{x_{-}}{W},\frac{y}{W}\right) -F_{2-}\left(\frac{x_{+}}{W},\frac{y}{W}\right)\right]\Bigg\}\nonumber\\
\end{eqnarray}
and
\begin{eqnarray}\label{eq:SolutionsDeviceGeometryVorticity}
\omega({\bm r}) &=& -\frac{2I}{e\bar{n}W^2}\Bigg[ F_{3-}\left(\frac{x_{-}}{W},\frac{y}{W},\frac{D_{\nu}}{W}\right)\nonumber\\
&-&F_{3-}\left(\frac{x_{+}}{W},\frac{y}{W},\frac{D_{\nu}}{W}\right)\Bigg]~.
\end{eqnarray}
In Eqs.~(\ref{eq:SolutionsDeviceGeometryPotential}) and~(\ref{eq:SolutionsDeviceGeometryVorticity}) we have introduce the shorthand $x_{\pm} \equiv x \pm x_0$, with $x_{+}$ ($x_{-}$) representing the lateral separation between the observation point and the collector (injector).

From Eq.~(\ref{eq:SolutionDecomposition}) we clearly notice an important feature of the solution, i.e.~for vanishing viscosity the current flow is irrotational.
More precisely, the viscosity plays a twofold role: it modifies the irrotational contribution due to the electric potential $\phi({\bm r})$ {\it and} introduces a finite vorticity.
It is noteworthy that these effects yield independent experimental signatures: the modification of the electrical potential can be detected by monitoring the resistances in a non-local configuration (or by carrying out scanning probe potentiometry), while the vorticity generates a magnetic field, which can be detected by scanning probe magnetometry. These two effects are discussed in detail in the following Sections.

\subsection{Spatial dependence of the 2D electrical potential, charge current, and non-local resistances}
\label{sect:potentiometry_and_rheometry}

\begin{figure}[t]
\begin{overpic}[width=\linewidth]{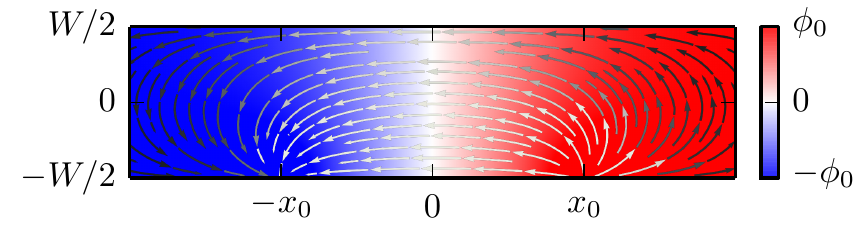}\put(2,65){(a)}\end{overpic}
\begin{overpic}[width=\linewidth]{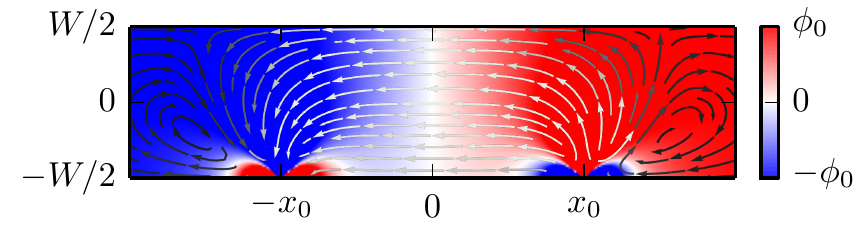}\put(2,65){(b)}\end{overpic}
\begin{overpic}[width=\linewidth]{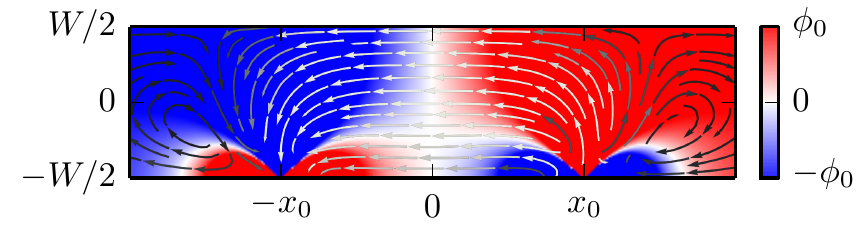}\put(2,65){(c)}\end{overpic}
\caption{\label{fig:nonlocal}
(Color online) Steady-state spatial map of the 2D electrical potential $\phi({\bm r})$ (in units of $\phi_0\equiv I/\sigma_0$) and charge current streamlines $-e\bm{J}({\bm r})$ in a Hall bar device like the one depicted in Fig.~\ref{fig:Device} with $x_0=W$.
Different panels refer to different values of the vorticity diffusion length $D_{\nu}$: $D_{\nu}=0$ [panel (a)], $D_{\nu}=0.5~W$ [panel (b)], and $D_{\nu} = W$ [panel (c)].
Whirlpools are clearly seen in the bottom right and bottom left of panels (b) and (c).
No whirlpools occur in the absence of viscosity, as in panel (a).
In each panel, the current streamlines change color from white (high current density) to black (low current density). }
\end{figure}

Illustrative results for the spatial map of the 2D electrical potential $\phi({\bm r})$---Eq.~(\ref{eq:SolutionsDeviceGeometryPotential})---and the charge current pattern $-e{\bm J}({\bm r})$---Eq.~(\ref{eq:SolutionDecomposition})---are shown in Fig.~\ref{fig:nonlocal}.
For typical values of the drive current $I$ and conductivity, i.e.~$I=20~{\rm \mu A}$ through a submicron constriction and $\sigma_0=20~{\rm mS}$, we find that the scale over which the 2D electrical potential changes is $\phi_0 \equiv I/\sigma_0 = 1~{\rm mV}$.
We clearly see that in the case $\nu \neq 0$---panels (b) and (c) in Fig.~\ref{fig:nonlocal}---whirlpools with a spatial extension $\sim D_{\nu}$ develop in the spatial current pattern $-e{\bm J}({\bm r})$, to the right of the current injector and to the left of the current collector.
Once again, the spatial variations of the 2D electrical potential $\phi({\bm r})$ are amenable to experimental studies based on scanning probe potentiometry.

In passing, we note that near the current injector at $x=x_0$ the potential is dominated by the singular parts of the functions $F_{m-}$.
Taking the limit $W/x_0\to \infty$ in Eq.~(\ref{eq:SolutionsDeviceGeometryPotential}) we find an extremely simple expression for the potential near the injector:
\begin{equation}\label{eq:pointinjector}
\phi(r,\theta)=-\frac{I}{\pi \sigma_0} \left[\ln \left( \frac{r}{R}\right)- 2 D^2_{\nu} \frac{\cos(2 \theta)}{r^2} \right]~,
\end{equation}
where $r$ is the distance from the injection point, $\theta$ is the angle measured from the injection direction $\hat{\bm y}$, and $R$ is a length determined by BCs far from the contact.
Note that changing $R$ is equivalent to changing $\phi$ by an arbitrary additive constant.

If dc transport is to be used as the main tool to detect hydrodynamic electron flow, it is pivotal to understand the spatial dependence of the non-local resistance $R_{\rm NL}$, which we define in the following way:
\begin{equation}\label{eq:NonLocalResistance}
R_{\rm NL}(x,y)\equiv \frac{\phi(x,y)-\phi(x\rightarrow +\infty,y)}{I}~,
\end{equation}
where the quantity $\phi(x\rightarrow +\infty,y)$ does not depend on $y$. Because of (\ref{eq:SolutionsDeviceGeometryPotential}), we find that, at each point in space, $R_{\rm NL}(x,y)$ is a quadratic function of $D_{\nu}$:
\begin{equation}\label{eq:LambdaDependence}
R_{\rm NL}(x,y)\sigma_0= a(x,y) D_{\nu}^2 + b(x,y)~,
\end{equation}
where
\begin{equation}\label{eq:a-expression}
a(x,y)=\frac{2}{W^2}\left[F_{2-}\left(\frac{x_{+}}{W},\frac{y}{W}\right)-F_{2-}\left(\frac{x_{-}}{W},\frac{y}{W}\right)\right]
\end{equation}
and
\begin{equation}\label{eq:b-expression}
b(x,y)= F_{1-}\left(\frac{x_{+}}{W},\frac{y}{W}\right)-F_{1-}\left(\frac{x_{-}}{W},\frac{y}{W}\right) - \frac{x_0}{W}~.
\end{equation}
To make contact with Ref.~\onlinecite{bandurin_arxiv_soon}, we now introduce the ``vicinity'' resistance, which is the non-local resistance measured on the edge where current is injected, at a distance $\Delta x$ from the current injector:
\begin{equation}\label{eq:vicinity_resistance}
R_{\rm V}(\Delta x) \equiv R_{\rm NL}(x_{0} + \Delta x, -W/2)~.
\end{equation}
Using Eqs.~(\ref{eq:LambdaDependence}), (\ref{eq:a-expression}), (\ref{eq:b-expression}), 
the asymptotic results in Table~\ref{tab:AsymptoticBehaviours}, and taking the limit $x_{0} \gg W$,  
we find
\begin{eqnarray}\label{eq:vicinity-expression}
R_{\rm V}(\Delta x) & = & \frac{-2 \pi  e^{-\pi |\Delta x|/W}}{W^2\left(1-e^{-\pi|\Delta x|/W}\right)^2}D_{\nu}^2+  \nonumber \\
& + & \left[-\frac{1}{\pi}\ln \left(1-e^{-\pi|\Delta x|/W}\right)+\Delta x~\Theta(-\Delta x)\right]~.\nonumber\\
\end{eqnarray}
Here $\Theta(x)$ is the Heaviside step function.
For positive $\Delta x$ the two terms in Eq.~(\ref{eq:vicinity-expression}) have opposite sign.
For this reason, $R_{\rm V}(\Delta x)$ is expected to change sign as a function of $D_{\nu}$. 
The change of sign of the vicinity resistance is a key signature of the viscous contribution to the electric potential. Maximum sensitivity to viscosity is achieved when slightly non-local or ``vicinity'' voltage drops are measured outside the region $[-x_0;x_0]$ where the current flux is maximum. The vicinity resistance (\ref{eq:vicinity-expression}) rapidly decays for $|\Delta x|\gg W/\pi$: it is therefore pivotal to measure~\cite{bandurin_arxiv_soon} the potential $\phi(x,y)$ for a lateral separation $\Delta x$ from the current injection point which is of the order of the vorticity diffusion length $D_{\nu}$.

Once the rate of momentum-non-conserving collisions $\tau^{-1}(\bar{n}, T)$ is measured from an ordinary four-point longitudinal measurement of $\sigma_{xx}$ (as explained in Sect.~\ref{sect:longitudinal}), a measurement of the vicinity resistance $R_{\rm V}(\Delta x)$ yields a map $\nu = \nu(\bar{n},T)$ of the kinematic viscosity of the 2D electron liquid. We hasten to stress that our all-electrical non-local protocol to measure the kinematic viscosity $\nu = \nu(\bar{n},T)$ applies to any 2D electron liquid driven into the hydrodynamic transport regime (and not only to doped SLG and BLG).

For the sake of completeness, we note that the authors of Ref.~\onlinecite{tomadin_prl_2014} have proposed an ac Corbino disk viscometer, which allows a determination of the hydrodynamic shear viscosity from the dc potential difference that arises between the inner and the outer edge of the disk in response to an {\it oscillating} magnetic flux.

\subsection{Spatial depedence of the current-induced magnetic field}

Because the steady-state current $-e {\bm J}({\bm r})$ generates a magnetic field in the proximity of 2D electron system, whirlpools and viscosity-dominated hydrodynamic transport can also be detected by scanning probe magnetometry (see, for example, Refs.~\onlinecite{taylor_naturephys_2008,maze_nature_2008,grinolds_naturephys_2013}).

As shown in Appendix~\ref{appendix:LIA}, in a sample in which a backgate is placed at a distance $z=-d$ below the graphene sheet with $d \ll W, D_{\nu}$, a local relation exists between the $\hat{\bm z}$-component $B_z({\bm r},z>0)$ of the magnetic field and the vorticity $\omega({\bm r})$. In SI units, this relation reads as following
\begin{equation}\label{eq:LIA}
B_z({\bm r},z>0) = -e\mu_0 \bar{n} d \omega({\bm r})~,
\end{equation}
where $\mu_0 = 4\pi \times 10^{-7}~{\rm N}/{\rm A}^2$ is the magnetic constant and the vorticity has been introduced earlier in Eq.~(\ref{eq:SolutionsDeviceGeometryVorticity}).

\begin{figure}
\begin{overpic}[width=\linewidth]{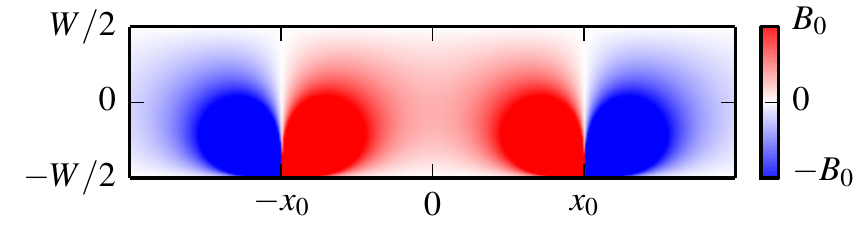}\put(2,65){(a)}\end{overpic}
\begin{overpic}[width=\linewidth]{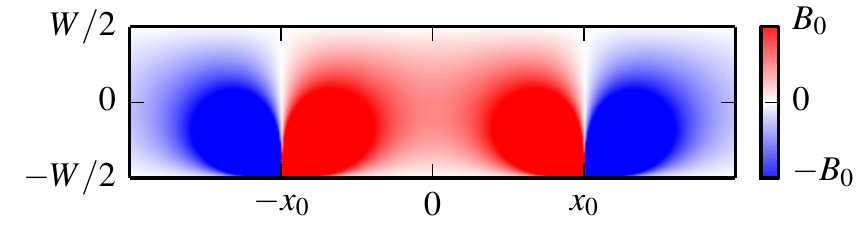}\put(2,65){(b)}\end{overpic}
\caption{(Color online) Spatial map of the $\hat{\bm z}$-component of the magnetic field, $B_z({\bm r},z)$ (in units of $B_0\equiv\mu_0 I d /W^2$), generated by the 2D steady-state current pattern ${\bm J}({\bm r})$ and calculated immediately above the graphene sheet, i.e.~for $0<z\ll W,D_{\nu}$.
These results have been obtained for the same parameters as in Fig.~\ref{fig:nonlocal}.
Different panels refer to different values of $D_{\nu}$: $D_{\nu}=0.5~W$ [panel~(a)] and $D_{\nu}=W$ [panel~(b)].
In this figure we have not shown results for $D_{\nu} = 0$: in the absence of viscosity $B_z({\bm r},z)$ is identically zero.\label{fig:magneticfield}}
\end{figure}

A typical 2D spatial map of $B_z({\bm r},z>0)$ is shown in Fig.~\ref{fig:magneticfield} for different values of the vorticity diffusion constant $D_{\nu}$.
In this figure the magnetic field is plotted in units of $B_0\equiv\mu_0 I d /W^2$.
For $I=200~{\rm \mu A}$, $W=1~{\rm \mu m}$, and $d=80~{\rm nm}$, we find $B_0=20~{\rm \mu T}$.
This value is well within reach of current technology~\cite{taylor_naturephys_2008,maze_nature_2008,grinolds_naturephys_2013}.

\section{Summary and future perspectives}
\label{sect:summary}

In summary, we have presented a theoretical study of dc transport in graphene driven into the hydrodynamic regime.
As highlighted in Ref.~\onlinecite{bandurin_arxiv_soon}, this regime occurs in a wide range of temperatures and carrier concentrations.

Our theory, which applies only to the doped regime, relies on the continuity (\ref{eq:ContinuityLinearised}) and Navier-Stokes (\ref{eq:NavierStokesLinearised_2}) equations, augmented by suitable boundary conditions at Hall bar edges.
We have demonstrated analytically that a combination of ordinary four-point longitudinal transport measurements and measurements of non-local resistances in Hall bar devices can be used to extract the hydrodynamic shear viscosity of the two-dimensional electron liquid in graphene~\cite{bandurin_arxiv_soon}.

We have also discussed how to probe the viscosity-dominated hydrodynamic transport regime by scanning probe methods.
Indeed, we believe that it possible to observe hydrodynamic electron flow with spatial resolution by using available scanning probe potentiometry and magnetometry setups.
Spatial maps of the two-dimensional electrical potential $\phi({\bm r})$ and current-induced magnetic field $B_z({\bm r}, z>0)$ for experimentally relevant parameters are shown in Figs.~\ref{fig:nonlocal} and~\ref{fig:magneticfield}.

We wish to emphasize that our theoretical approach is immediately applicable to any 2D electron liquid in the hydrodynamic transport regime.

In the future, we plan to extend our theoretical approach to semimetals with linear and quadratic band touchings at the charge neutrality point, where viscosity is expected to be very low~\cite{muller_prl_2009} and the Reynolds number (\ref{eq:reynoldsnumber}) is expected to be very large, possibly enabling the observation of electronic turbulence.
This will require to deal with thermally-excited carriers, coupling between charge and heat flow, and non-linear terms in the Navier-Stokes equation.
We also would like to gain a fully microscopic understanding of momentum-non-conserving collisions in the hydrodynamic transport regime by treating smooth scalar {\it and} vector potentials due to disorder (strain~\cite{gibertini_prb_2012}, charged impurities~\cite{dassarma_rmp_2011}, etc.) along the lines of what was done by the authors of Ref.~\onlinecite{andreev_prl_2011} for the case of a smooth scalar potential.

Last but not least, we strongly believe that hydrodynamic flow and the shear viscosity of 2D electron liquids can also be accessed~\cite{torre_prb_2015} by scattering-type near-field optical spectroscopy (see, for example, Ref.~\onlinecite{woessner_naturemater_2015} and references therein to earlier work) in the Terahertz spectral range, since this technique measures the non-local conductivity $\sigma(q,\omega)$. Terahertz radiation is required a) to make sure that the hydrodynamic inequality $\omega \tau_{\rm ee} \ll 1$ is satisfied (with $\tau_{\rm ee} = \ell_{\rm ee}/v_{\rm F}$) and b) to have measurable non-local effects due to viscosity, since the latter decreases quickly~\cite{principi_arxiv_2015} as a function of the external probe frequency $\omega$.

\appendix

\section{Local inductance approximation}
\label{appendix:LIA}

In this Appendix we present a derivation of Eq.~(\ref{eq:LIA}).
We use SI units and the Coulomb gauge $\nabla \cdot \bm{A}=0$ for the vector potential.
We assume that a bottom gate positioned at $z=-d$ exists below the graphene sheet.
This will be treated as a perfect conductor.

The vector potential is related to the steady-state current pattern by the following 3D Poisson equation:
\begin{equation}\label{eq:Poissonvector}
\left(\nabla^2 +\frac{\partial^2}{\partial z^2}\right)\bm{A}(\bm{r},z)=\mu_0 e \bm{J}(\bm{r})\delta(z)~.
\end{equation}
This is similar to the Poisson equation in Eq.~(\ref{eq:poisson}) for the scalar potential $\Phi({\bm r},z)$.

Fourier transforming Eq.~(\ref{eq:Poissonvector}) with respect to the in-plane coordinate ${\bm r}$ we find:
\begin{equation}\label{eq:poissonvectorq}
\left(-q^2 +\frac{\partial^2}{\partial z^2}\right)\ft{\bm{A}}(\bm{q},z)=\mu_0 e \ft{\bm{J}}(\bm{q})\delta(z)~,
\end{equation}
where ${\bm q} \cdot \ft{\bm{J}}(\bm{q})=0$ because of the continuity equation Eq.~(\ref{eq:ContinuityLinearised}).

The general solution of Eq.~(\ref{eq:poissonvectorq}) is:
\begin{equation}
\ft{\bm{A}}(\bm{q},z)=-\mu_0 e \ft{\bm{J}}(\bm{q})\frac{e^{-q|z|}}{2q} + \ft{\bm{a}}_+(\bm{q})e^{qz} +
\ft{\bm{a}}_-(\bm{q})e^{-qz}~.
\end{equation}
The quantities $\ft{\bm{a}}_{\pm}(\bm{q})$ must obey the condition $\ft{a}_{\pm z}(\bm{q})=\mp i \bm{q} \cdot \ft{\bm{a}}_{\pm }(\bm{q})/q$ to enforce the Coulomb gauge and must be determined from BCs.
Requiring $\ft{\bm{A}}(\bm{q},z \to +\infty)=0$ implies that $\ft{\bm{a}}_+(\bm q)$ must vanish.

The corresponding $z$ component of the magnetic field is given by:
\begin{equation}\label{eq:Bz}
\ft{B}_z({\bm q}, z)=[i \bm{q} \times \ft{\bm{A}}(\bm{q},z)]_z~.
\end{equation}
Since $B_z$ must vanish at the gate position, i.e.~for $z=-d$,
we find: $\ft{\bm a}_-(\bm{q})=\mu_0e\ft{\bm J}({\bm q})e^{-2qd}/(2q)$.
In deriving the previous result we have assumed that $\ft{\bm a}_{-}({\bm q})\cdot \bm{q}=0$.

The Fourier transform of the vector potential is therefore given by:
\begin{equation}
\ft{\bm A}({\bm q}, z) = -\mu_0 e \ft{\bm J}({\bm q})\frac{e^{-q|z|}-e^{-q(2d+z)}}{2q}~.
\end{equation}
Now, if $d$ and $|z|$ are small with respect to the lateral lengthscales $W$ and $D_{\nu}$ over which the steady-state current pattern ${\bm J}({\bm r})$ changes in the sample, i.e.~$d,|z|\ll W,~D_{\nu}$, the above formula can be approximated for $z>0$ as:
\begin{equation}
\ft{\bm{A}}(\bm{q},z>0) \approx -e \mu_0 d \ft{\bm{J}}(\bm{q})~.
\end{equation}
Making use of Eq.~(\ref{eq:Bz}) and transforming back to real space we finally obtain the desired result,
\begin{equation}
B_z(\bm{r},z>0)\approx -e \mu_0 d[\nabla \times \bm{J}(\bm{r})]_z = -e \mu_0 \bar{n}d \omega(\bm{r})~.
\end{equation}

\acknowledgements
This work was partially supported by MIUR through the program ``Progetti Premiali 2012'' - Project ``ABNANOTECH''. Free software (www.gnu.org, www.python.org) was used. 
We thank M.F.~Crommie, L.S.~Levitov, L.A.~Ponomarenko, G. Vignale, and A.~Yacoby for fruitful discussions.

\end{document}